\documentclass[aps,twocolumn,prd,showpacs,nofootinbib]{revtex4}

\usepackage{amsmath}
\usepackage{graphicx}
\usepackage{dcolumn}
\usepackage{bm}
\usepackage{amssymb}
\usepackage{latexsym}

\def\setC{\mathbb{C}}

\def\setR{\mathbb{R}}

\newcommand{\gsim}{\gtrsim}

\newcommand{\ie}{\textsl{i.e.~}}

\newcommand{\etal}{\textsl{et al.~}}

\newcommand{\mP}{m_{_{\mathrm Pl}}}
\newcommand{\GReCO}{${\cal G}\setR\varepsilon\setC{\cal O}$}

\def\spose#1{\hbox to 0pt{#1\hss}}
\def\lta{\mathrel{\spose{\lower 3pt\hbox{$\mathchar"218$}}
     \raise 2.0pt\hbox{$\mathchar"13C$}}}
\def\gta{\mathrel{\spose{\lower 3pt\hbox{$\mathchar"218$}}
     \raise 2.0pt\hbox{$\mathchar"13E$}}}

\begin{document}

\title{Coupling Quintessence to Inflation in Supergravity}

\author{Philippe Brax \footnote{Electronic mail:
    brax@spht.saclay.cea.fr \\ Associate Researcher
at  Institut d'Astrophysique de Paris, \GReCO, FRE 2435-CNRS,
98bis boulevard Arago, 75014 Paris, France}}
\affiliation{Service de Physique Th\'eorique, CEA-Saclay,
Gif/Yvette cedex, France F-91191}

\author{J\'er\^ome Martin \footnote{Electronic mail: jmartin@iap.fr}}
 \affiliation{Institut d'Astrophysique de
Paris, \GReCO, FRE 2435-CNRS, 98bis boulevard Arago, 75014 Paris,
France}

\date{\today}

\begin{abstract}
The evolution of the quintessence field during a phase of chaotic
inflation is studied. The inflaton $\phi$ and the quintesssence field
$Q$ are described in a supergravity framework where the coupling
between the inflaton and quintessence is induced by non-renormalisable
operators suppressed by the Planck mass. We show that the resulting
quintessence potential during inflation possesses a time--dependent
minimum playing the role of an attractor. The presence of this
attractor forces the quintessence field to be small during
inflation. These initial conditions are such that the quintessence
field is on tracks now.
\end{abstract}

\maketitle
\section{Introduction}

A host of recent cosmological observations, the anisotropy of the
Cosmic Microwave Background (CMB)~\cite{wmap,ISW}, the large scale
structures of the Universe~\cite{sdss} and type Ia
supernovae~\cite{SNIa}, indicate that the Universe has experienced two
stages of cosmic acceleration. The first one is the inflationary era
which occurred in the early Universe~\cite{inflation}. It is
responsible for the almost flatness of the Universe and primordial
density fluctuations~\cite{pertinflation} (see also
Refs.~\cite{MS}). The second one, which started in the recent past,
leads to the present acceleration of the expansion of the
Universe. Various explanations for this last phenomenon have been
proposed in the literature: a pure cosmological constant~\cite{Wein},
quintessence~\cite{RP,quint,PB,BM1,BM2,BMR1,BMR2},
k-essence~\cite{kessence}, modified gravity theories~\cite{cedric},
the Chaplygin gas~\cite{julio}, bulk viscosity~\cite{dominik} or
quantum cosmological effects~\cite{nelson}. In this paper, we focus on
the quintessence hypothesis. In this case, the two phenomena described
above are modeled as resulting from the presence of two scalar fields
whose energy densities drive the acceleration of the expansion. The
quintessence hypothesis has been further investigated in
Refs.~\cite{quintfurther}. In particular, finding a natural candidate
for the quintessence field in the realm of high energy physics has
been a major goal for lot of authors~\cite{BM1,BM2,BMR2,modelquint} as
well as studying some aspects of its interaction with the ``rest of
the world''~\cite{restworld}. Since, contrary to a cosmological
constant, the quintessence field can develop some inhomogeneities, the
theory of cosmological perturbations has also been studied in
details~\cite{BMR1,quintpert} and has been used in order to constraint
various models observationally ~\cite{observfurther}.  Of course, the
prospect of utilizing the fact that the quintessence equation of state
is no longer time (or redshift)-independent as a tool for
discriminating amongst the various possibilities has been widely
discussed in the recent literature~\cite{paramstate}.

\par

As a low energy description of string theory~\cite{polchinski},
supergravity captures prominent features of physics beyond the
standard models of particle physics and cosmology. Supergravity is the
best framework within which both quintessence and inflation can be
described. Indeed inflation (in its most common models like chaotic
inflation) involves high energies as the inflaton rolls down its
potential with values exceeding the Planck mass.  Similarly, in
quintessence models with a rolling scalar, the quintessence field
reaches values of the order of the Planck mass now. Hence the
necessity for a treatment where non-renormalizable interaction terms
suppressed by the Planck mass are under control.  In supergravity,
such non-renormalizable corrections to supersymmetric models are taken
into account and play an important role. This justifies the use of
supergravity models both in inflation and quintessence model
building. In the following we will concentrate on both quintessence
and inflation as described in supergravity.

\par

One of the commonly used models of quintessence, first devised by
Ratra and Peebles~\cite{RP}, requires an inverse power law behavior
$V(Q)=M^{4+\alpha}Q^{-\alpha}$ with an attractor mechanism at large
time. It was soon realized that this type of potential can be
generated in supersymmetric theories when a strongly interacting
sector is present~\cite{PB}. In particular, the value of the
quintessence field becomes of the order of the Planck scale which
prompts the necessity of a supergravity treatment. A simple embedding
of the previous model in supergravity fails as the potential is highly
modified by supergravity corrections and can become
negative~\cite{BM1,BM2}.  Hence a more phenomenological approach may
be required where one postulates the form of the K\"ahler potential
and the superpotential which leads to quintessence in
supergravity. This was done in Refs.~\cite{BM1,BM2} and subsequent
work.

\par

Similarly, as a high energy phenomenon occurring in the early
universe, inflation must be described within supergravity.  Recently,
there has been a upsurge of inflation models in supergravity motivated
by string theory~\cite{kallosh,maldacena,quevedo}. It seems natural to
study the influence of quintessence on inflation and vice versa.

\par

Quintessence must be almost decoupled from ordinary matter,
otherwise the quintessence field would lead to observable fifth
force signals~\cite{restworld}. On the contrary the inflaton field
must couple quite strongly to ordinary matter in order to have a
reheating period at the end of inflation where the oscillations of
the inflaton result in a radiation bath. Hence the coupling of
quintessence and inflaton cannot be large. A natural way of
realizing this criterion is to consider that the inflaton and the
quintessence field are decoupled in the K\"ahler potential and the
superpotential of supergravity. This implies that the only
possible interactions between both fields spring from
non-renormalizable interactions suppressed by the Planck mass.
Here we provide such a supergravity description of the coupling
between inflation and quintessence.

\par

The paper is arranged as follows. In a first part (Sec.~II), we
analyze the supergravity coupling between a particular quintessence
model, the so-called SUGRA model~\cite{BM1,BM2,BMR2}, and a generic
inflationary model. We then (Sec.~III) apply this analysis to the
specific example of chaotic inflation where we show that the
quintessence field develops a potential with a rolling minimum during
inflation. The rolling minimum is an attractor such that the values of
the quintessence field remain small throughout the inflationary
era. In particular, these values are much smaller than the values of a
free quintessence field during inflation. The smallness of the
quintessence field during inflation implies that it is on tracks now,
i.e. it reaches its long time attractor. Finally, in Sec.~IV, we
discuss the limitations of our approach, try to indicate what possible
improvements could be and present our conclusions.

\section{Quintessence and Inflation in Supergravity}

\subsection{Quintessence in Supergravity}

Let us now briefly review a simple model of quintessence in
supergravity often dubbed the SUGRA model in the literature. We assume
that the K\"ahler potential and the superpotential are given
by~\cite{BM1,BM2}
\begin{eqnarray}
K_{\rm quint}\left(X,Y,Q\right) &=& XX^{\dagger }+QQ^{\dagger} +\kappa
^pYY^{\dagger} \left(QQ^{\dagger}\right)^p\, ,
\\
W_{\rm quint}\left(X,Y,Q\right) &=& \mu X^2Y\, ,
\end{eqnarray}
with $\kappa \equiv 8\pi/\mP^2$. Here $X$ and $Y$ are two charged
fields under an (anomalous) $U(1)$ symmetry with charges $1$ and $-2$,
while $Q$ is the neutral quintessence field.  Notice the direct
coupling between $Q$ and $Y$. The constant $\mu$ is a dimensionless
coupling constant and $p$ is a free coefficient. It is worth
mentioning that one can derive the SUGRA model from more general
K\"ahler potentials but we will not need them in this article, see
Ref.~\cite{BMR2}. At this stage, we assume that
\begin{equation}
\langle X \rangle= \xi \, ,\quad \langle Y \rangle =0\, .
\end{equation}
As a specific example, $\xi$ can be realized as a Fayet-Iloupoulous
term arising from the Green--Schwarz anomaly cancellation mechanism.
When $\mu =0$, the $Q$ direction is flat. It is lifted by the
superpotential leading to the quintessence potential. In supergravity,
negative contributions to the scalar potential arise from the vacuum
expectation value (vev) of the superpotential.  Notice that we have
here
\begin{equation}
\left \langle W_{\rm quint} \right \rangle =0\, ,
\end{equation}
implying that no negative contribution appears in the scalar
potential.

\par

We are now in a position where the scalar potential can be
computed. In supergravity, it is given by
\begin{equation}
V=\frac{1}{\kappa ^2}{\rm e}^G\left(G^AG_A-3\right)\, ,
\end{equation}
where the matrix $G_{A\bar{B}}$ which is used to raise and lower the
index $A$ is defined by
\begin{equation}
G_{A\bar{B}}=\frac{\partial ^2}{\partial \varphi ^A \partial
\left(\varphi ^B\right)^{\dagger}} \left[\kappa K_{\rm quint}+\ln
\left(\kappa ^3\left\vert W_{\rm quint} \right\vert^2\right)\right]\,
,
\end{equation}
where $\varphi ^A=\{X,Y,Q\}$ are the fields in the quintessence
sector. Straightforward calculations leads to a matrix which is
block--diagonal, namely
\begin{equation}
G_{A\bar{B}}=\kappa \left[
\begin{matrix}
1 & 0 & 0 \cr
0 & \left(\kappa QQ^{\dagger }\right)^p & 0 \cr
0 & 0 & 1
\end{matrix}\right]\, .
\end{equation}
Then the complete SUGRA potential becomes
\begin{equation}
\label{vquint}
V_{\rm quint}(Q)={\rm e}^{\kappa Q^2+\kappa \xi
^2}\frac{M^{4+2p}}{Q^{2p}}\, ,
\end{equation}
where the mass scale $M$ characterizing the potential can be
expressed as $M^{4+2p}\equiv \mu ^2\xi ^4 \kappa ^{-p}$. It is
easy to find that $\left(K_{\rm quint}\right)_{QQ^{\dagger }}=1$
which means that the real part field $Q$ is in fact not correctly
normalized. Therefore, one has to redefine the field $Q$ according
to $Q\to Q/\sqrt{2}$ and this gives
\begin{equation}
V_{\rm quint}(Q)={\rm e}^{\kappa Q^2/2+\kappa \xi
^2}\frac{M^{4+2p}}{Q^{2p}}\, ,
\end{equation}
where we have slightly redefined the mass scale $M$ such that
$M^{4+2p} \to M^{4+2p}\times 2^p$. The main feature of the above
potential is that supergravity corrections have been exponentiated and
appear in the prefactor. Phenomenologically, this potential has the
nice feature that the equation of state $\omega \equiv p_Q/\rho _Q$
can be closer to $-1$ than with the Ratra--Peebles potential.

\subsection{Inflation in Supergravity}

Let us now give a brief description of the inflation models we will
concentrate on. We will consider a class of models described by the
following K\"ahler potential
\begin{eqnarray}
\label{kahlerinf}
K_{\rm inf} &=& -\frac{3}{\kappa }\ln \left[\kappa ^{1/2}\left(\rho+ \rho
^{\dagger }\right)-\kappa {\cal K}\left(\phi-\phi^\dagger
\right)\right]
\nonumber \\
& & +{\cal G}\left(\phi-\phi^\dagger\right)\, ,
\end{eqnarray}
where ${\cal K}$ and ${\cal G}$ are arbitrary functions. The field
$\phi $ is the inflaton while $\rho $ represents, for instance, a
moduli of a string compactification. The superpotential $W_{\rm
inf}=W_{\rm inf}\left(\rho ,\phi \right)$ is not specified at this
stage. This form is justified by the fact that one can obtain flat
enough potentials in supergravity by requiring that a shift symmetry
$\phi\to \phi +c$, where $c$ is a real constant, is a symmetry of the
K\"ahler potential, later broken mildly. The K\"ahler potential given
by Eq.~(\ref{kahlerinf}) obviously possesses this symmetry. Indeed, a
striking feature of F-term inflation in supergravity is the natural
presence of ${\cal O}(H_{\rm inf})$ corrections to the inflaton mass
which would spoil the flatness of the potential. These problems can be
avoided by considering the above class of models.

\par

To go further, one must specify the functions ${\cal K}$, ${\cal G}$
and the superpotential. We choose an example of chaotic inflation as
can be found in Ref.~\cite{linde} where a similar case is
treated. Explicitly, one assumes
\begin{eqnarray}
{\cal K} &=& -\frac{1}{2}\left(\phi -\phi ^{\dagger }\right)^2\, , \quad
{\cal G}=+\frac{1}{2}\left(\phi -\phi ^{\dagger }\right)^2\, ,
\end{eqnarray}
and for the superpotential
\begin{eqnarray}
W_{\rm inf}(\rho, \phi ) &=& \frac{\alpha }{2}m\phi^2 \, .
\end{eqnarray}
The factor $\alpha $ in the superpotential is free and can be chosen
for future convenience. Notice that the shift symmetry is preserved by
our choice of the functions ${\cal K}$ and ${\cal G}$ while, on the
contrary, the superpotential breaks this symmetry explicitly. Then,
straightforward calculations lead to
\begin{equation}
V_{\rm inf}(\rho ,\phi)=\frac{1}{\Delta ^2(3-\Delta )}\alpha ^2m^2\phi
^2\, ,
\end{equation}
where $\Delta =\kappa^{1/2}(\rho +\rho ^\dagger )$. It is easy to see
that the moduli can be stabilized if $\Delta =2$. Furthermore, one can
check that the normalization of the inflaton is given by $\left(K_{\rm
inf}\right)_{\phi \phi^\dagger }=3/\Delta -1=1/2$ and, hence, is
correct. In this case, the potential takes the form
\begin{equation}
V_{\rm inf}(\phi)=\frac{\alpha ^2}{4} m^2\phi^2 \, ,
\end{equation}
which is nothing but the usual chaotic inflation potential if one
chooses $\alpha =\sqrt{2}$.

\par

Of course, it is possible to discuss more complicated and/or general
inflationary models. The one considered here has the advantage to lead
to the prototypical single field inflationary model, namely chaotic
inflation. Since our main goal is not to study inflation itself but
the coupling of the inflaton with the quintessence field, this model
is sufficient. However, it is clear that the next step would be to
study how the form of the coupling term that we are going to derive
below depends on the assumed inflationary model.

\subsection{Coupling the Inflaton to the Quintessence Field}

We now turn to our main goal, namely the calculation of the coupling
between the inflaton field and the quintessence field. Our basic
assumption is that the quintessence and inflation sectors are
decoupled, {\it i.e.}  that the total K\"ahler potential and
superpotential can be written as
\begin{eqnarray}
K &=& K_{\rm quint}\left(X,Y,Q\right)+K_{\rm inf}\left(\rho ,\phi
\right)\, , \\ W &=& W_{\rm quint}\left(X,Y,Q\right)+W_{\rm
inf}\left(\rho ,\phi\right)\, ,
\end{eqnarray}
where the quintessential K\"ahler potential and superpotential have
been given before but where, at least at this stage, the inflationary
part is still arbitrary. However, later , we will restrict our
considerations to the (chaotic) inflation model studied in the
preceding section. From the above equations, one deduces that the
matrix $G_{A\bar{B}}$, where now $\varphi ^A=\{X,Y,Q,\rho, \phi \}$,
is diagonal by blocks. Explicitly, one has
\begin{equation}
G_{A\bar{B}}=\left[
\begin{matrix}
G_{\rho \rho^{\dagger}} & G_{\rho \phi ^{\dagger }} & 0 & 0 & 0 \cr
G_{\phi \rho ^{\dagger}} & G_{\phi \phi ^{\dagger }} & 0 & 0 &0 \cr
0 & 0 & \kappa & 0 & 0 \cr
0 & 0 & 0 & \kappa \left(\kappa QQ^{\dagger }\right)^p & 0 \cr
0 & 0 & 0 & 0 & \kappa
\end{matrix}\right]\, .
\end{equation}
Then, the scalar potential takes the form (recall that the D-terms
contribution vanishes)
\begin{widetext}
\begin{equation}
V={\rm e}^{\kappa \xi ^2}\left[{\rm e}^{\kappa Q^2/2}V_{\rm
inf}\left(\rho, \phi \right)+{\rm e}^{\kappa K_{\rm inf}}V_{\rm
quint}\left(Q\right) +\kappa ^2\left(\xi
^2+\frac{Q^2}{2}\right)\left\vert W_{\rm inf}\right \vert ^2 {\rm
e}^{\kappa \left(K_{\rm inf}+Q^2/2\right)}\right]\, ,
\end{equation}
\end{widetext}
where,
\begin{eqnarray}
V_{\rm inf}\left(\rho ,\phi \right) &=& \frac{1}{\kappa ^2}
{\rm e}^{G_{\rm inf}}\left[G_{\rm
inf}^A\left(G_{\rm inf}\right)_A-3\right]\, ,
\\
V_{\rm quint}\left(Q\right) &=& {\rm e}^{\kappa Q^2/2}
\frac{M^{4+2p}}{Q^{2p}}\, .
\end{eqnarray}
Let us notice that, for convenience, we have slightly changed the
notation for $V_{\rm quint}(Q)$. Now, we no longer include the factor
$\exp(\kappa \xi ^2)$ in its definition, see Eq.~(\ref{vquint}). As
explained before, we have also redefined the quintessence field
according to $Q\to Q/\sqrt{2}$ in order to work with correctly
normalized fields. The above expression represents the general form of
the coupling between the SUGRA model of quintessence and inflation in
supergravity.

\par

\begin{figure*}
\includegraphics[width=.85\textwidth,height=.65\textwidth]{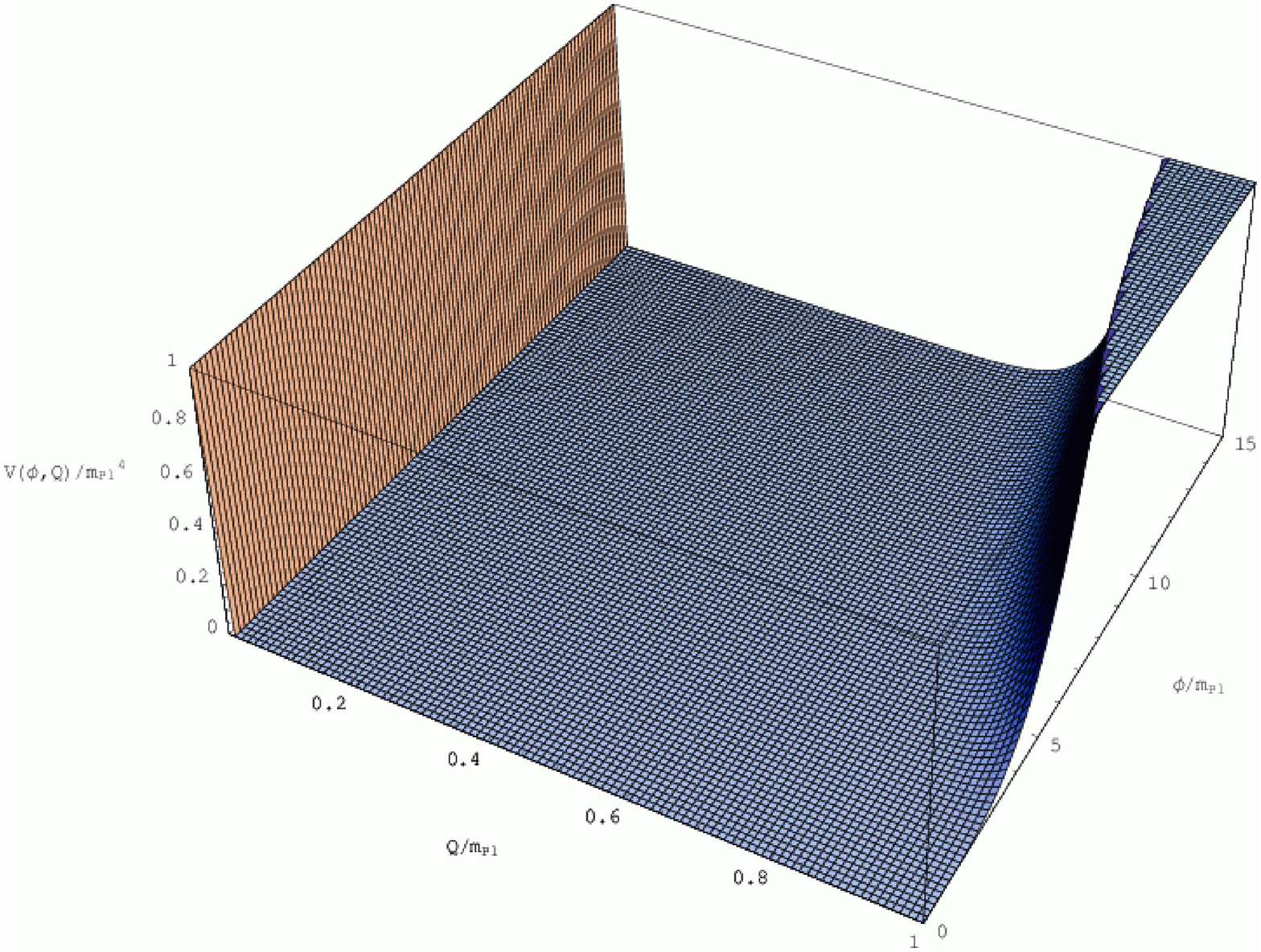}
\includegraphics[width=.85\textwidth,height=.65\textwidth]{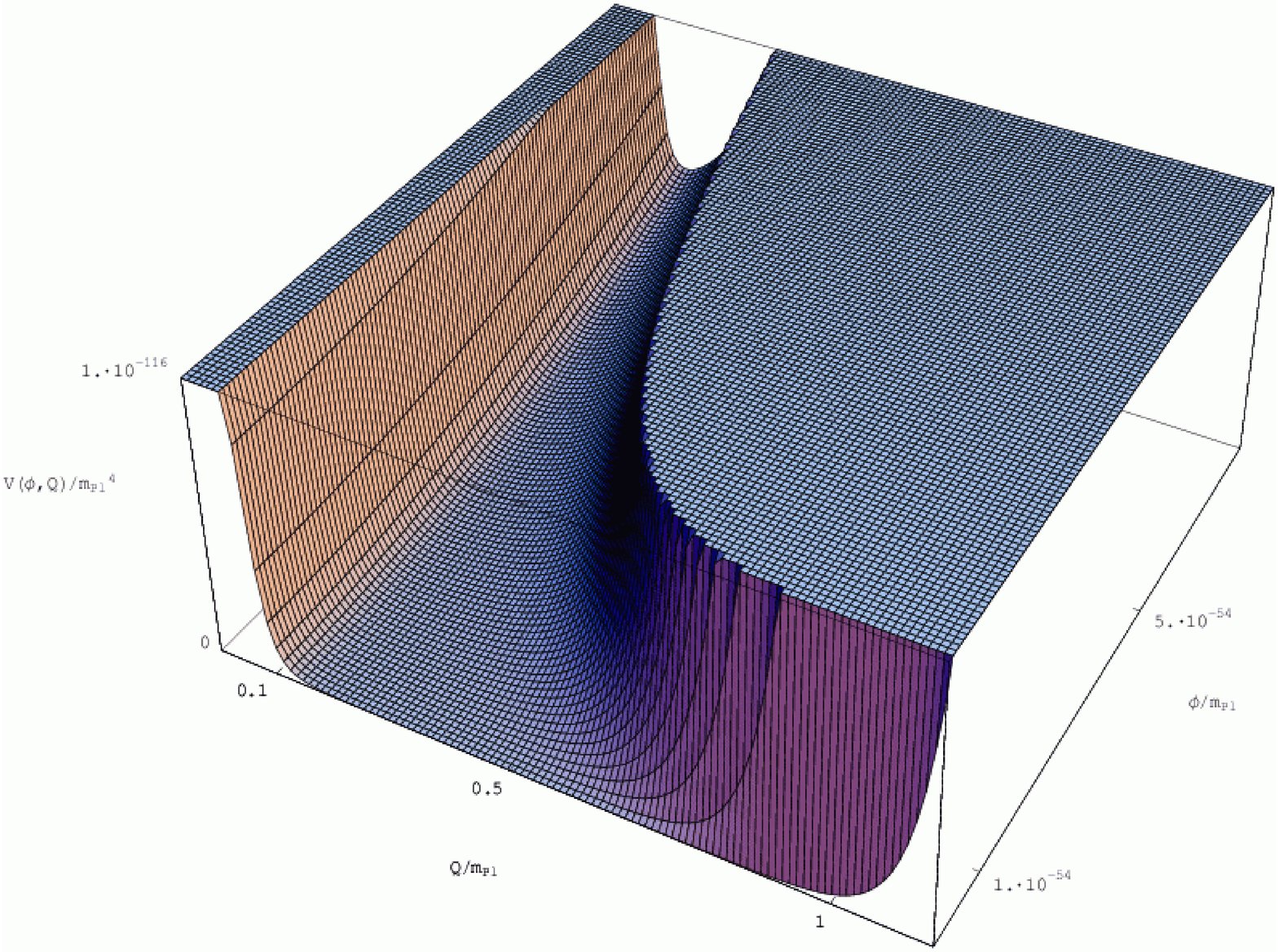}
\caption{Upper panel: Potential $V(\phi ,Q)$ for the following choice
of parameters: $p=3$, $m=10^{-5}\mP$, $\xi =10^{-30}\mP$, and
$(M/\mP)^{4+2p}=10^{-122}$. The absolute minimum located at $Q\simeq
0.4886 \times \mP$ and $\phi =0$ cannot be viewed with the scales
used. Bottom panel: zoom in the region of the potential where the
minimum is located. It is clear that the tiny values of the inflaton
field are, in this panel, not interesting from a physical point of
view (there is no inflation for such small values).}
\label{potqphi}
\end{figure*}

We now specify the inflaton model and consider the model described in
the previous subsection with $\alpha =\sqrt{2}$. It is convenient to
work in terms of dimensionless quantities. In particular, we define
the dimensionless potential $f\left(\phi ,Q\right)$ by $V(\phi
,Q)\equiv \mP^4 f(\phi ,Q)$ and this quantity can be written as
\begin{equation}
\label{potqphi1}
f\left(\phi ,Q \right)=f_{\rm inf}+f_{\rm quint}+f_{\rm inter}\, ,
\end{equation}
where
\begin{widetext}
\begin{eqnarray}
\label{potqphi2}
f_{\rm inf} &=& \frac12\left(\frac{m}{\mP}\right)^2\left(\frac{\phi
}{\mP}\right)^2{\rm e}^{8\pi \xi ^2/\mP^2+4\pi Q^2/\mP^2}\, , \quad
f_{\rm quint} = \frac18
\left(\frac{M}{\mP}\right)^{4+2p}\left(\frac{Q}{\mP}\right)^{-2p} {\rm
e}^{8\pi \xi ^2/\mP^2+4\pi Q^2/\mP^2}\, , \nonumber \\ f_{\rm inter}
&=& 4\pi ^2\left(\frac{m}{m_{_{\rm Pl}}}\right)^2
\left[\left(\frac{\xi
}{\mP}\right)^2+\frac12\left(\frac{Q}{\mP}\right)^2
\right]\left(\frac{\phi }{\mP }\right)^4 {\rm e}^{8\pi \xi
^2/\mP^2+4\pi Q^2/\mP^2}\, .
\end{eqnarray}
At this point, some remarks are in order. A priori, there is no clear
separation between the inflaton and the quintessence fields in the
term $f_{\rm inf}$ because of the presence of the exponential
term. However, in the regime we will be studying (during inflation),
$Q\ll \mP$ and, therefore, the exponential term will be very close to
one. In this case, one recovers the simple chaotic model $V_{\rm
inf}=m^2\phi ^2/2$. The term $f_{\rm quint}$ is nothing but the SUGRA
potential studied in Ref.~\cite{BM1,BM1,BMR2} but, during inflation,
it will reduce to the Ratra-Peebles case. Let us notice that we have
an extra factor $1/8$ originating from the term $\exp\left(\kappa
K_{\rm inf}\right)$. This comes from the fact that $\kappa K_{\rm
inf}=-3\ln (2)$ since the moduli is stabilized at $\Delta =2$.
Finally, in the regime $Q\ll \mP$, the interaction term reads $V_{\rm
inter}\propto m^2\phi ^4Q^2/\mP^4$. This is due to the fact that we
have $Q\gg \xi $ as will be discussed below. The coupling constant
between the inflaton and the quintessence fields reads
$m^2/\mP^4$. The Planck mass appears in this expression because the
coupling between the two fields has been entirely fixed by the
supergravity. Notice also that the quintessence field picks up an
inflaton dependent mass term during inflation. The competition between
this mass term and the Ratra--Peebles potential will be studied in the
next section. Finally, using the fact that
\begin{eqnarray}
\frac{\partial f}{\partial (\phi /\mP)} &=&
{\rm e}^{8\pi \xi ^2/\mP^2+4\pi Q^2/\mP^2}
\left(\frac{m}{m_{_{\rm Pl}}}\right)^2
\frac{\phi}{\mP}\left\{1+16\pi ^2\left[\left(\frac{\xi
}{\mP}\right)^2+\frac12\left(\frac{Q}{\mP}\right)^2\right]
\left(\frac{\phi }{\mP}\right)^2\right\}\, ,
\\
\frac{\partial f}{\partial (Q/\mP)} &=&
{\rm e}^{8\pi \xi ^2/\mP^2+4\pi Q^2/\mP^2}
\frac18\left(\frac{M}{\mP}\right)^{4+2p}
\left(\frac{Q}{\mP}\right)^{-2p}
\left[8\pi \frac{Q}{\mP}-2p\left(\frac{Q}{\mP}\right)^{-1}\right]
\nonumber \\
&+& {\rm e}^{8\pi \xi ^2/\mP^2+4\pi Q^2/\mP^2}
\left(\frac{m}{\mP}\right)^2
\left(\frac{\phi }{\mP}\right)^2
\frac{Q}{\mP}
\biggl\{ 4\pi +4\pi ^2\left[1+8\pi
\left(\frac{\xi }{\mP }\right)^2
+4\pi \left(\frac{Q}{\mP}\right)^2\right]\left(\frac{\phi }{\mP}\right)^2
\biggr\}\, ,
\end{eqnarray}
\end{widetext}
it is easy to show that this potential possesses an absolute minimum
given by
\begin{equation}
\frac{\phi }{\mP}=0 \, ,\quad \frac{Q}{\mP}=\sqrt{\frac{2p}{8\pi }}\,
.
\end{equation}
The potential is represented in Fig.~\ref{potqphi}. The value $\phi
=0$ is the minimum of the inflaton potential without interaction while
$Q=\sqrt{p/(4\pi )}$ is the minimum of the SUGRA potential. At the
absolute minimum, the value of the potential is non-vanishing and
given by $V= V_{\rm quint}=(\mP ^4/8) (M/\mP)^{4+2p}(4\pi /p)^p
\exp(\kappa \xi ^2)\exp (p)$.

\section{Cosmological Evolution}

\subsection{Fixing the free parameters}

Let us now discuss the values of the free parameters that appear in
Eqs.~(\ref{potqphi1}) and~(\ref{potqphi2}). If we assume that the
quintessential part is responsible for the acceleration now then one
should have
\begin{equation}
{\rm e}^{\kappa \xi ^2+\kappa Q_0^2/2}\frac{M^{4+2p}}{Q^{2p}_0}\simeq
m_{_{\rm Pl}}^2H_0^2\, ,
\end{equation}
where $Q_0$ and $H_0$ denote the values of the quintessence field and
of the Hubble parameter (respectively) now, at vanishing redshift. In
order to have a successful model of quintessence, the field should be
on track today which in turn implies that $Q_0= {\cal
O}\left(\mP\right)$. Strictly speaking, this conclusion is valid for
the Ratra-Peebles potential only, but it has been shown in
Ref.~\cite{BM1,BM2} that this is also valid for the SUGRA potential
despite the presence of the exponential correction (this is simply
because, except at small redshifts, we have $Q\ll \mP $ and the
exponential SUGRA correction does not play an important role). This
gives
\begin{equation}
\left(\frac{M}{\mP}\right)^{4+2p}\simeq
\frac{H_0^2}{m_{_{\rm Pl}}^2}\simeq 10^{-122}\, .
\end{equation}
Using the fact that $M^{4+2p}\simeq \mu ^2\xi ^4\kappa ^{-p}$ and
assuming no fine-tuning of the coupling constant, \ie $\mu ={\cal
O}\left(1\right)$, one deduces that
\begin{equation}
\frac{\xi }{m_{_{\rm Pl}}}\simeq \sqrt{\frac{H_0}{m_{_{\rm Pl}}}}
\simeq 10^{-30}\, .
\end{equation}
In a sense this is the usual fine--tuning of the cosmological
constant, it reappears here in the guise of the tuning of the vev of a
field leading to the quintessence potential. However, if one works
with an effective model valid up to a cut-off scale $m_{_{\rm C}}$,
then it has been shown in Ref.~\cite{BMR2} that the previous problem
can be solved provided the scale is chosen such that $m_{_{\rm C}}\ll
m_{_{\rm Pl}}$ . Then the value of the Fayet--Iliopoulos term can even
be above the weak scale. However, again, our purpose here is not to
study the details of the dark energy model and, therefore, in the
following, we will ignore these subtleties and work with the value of
$\xi $ derived before.  Let us notice that even with a
Fayet--Iliopoulos term above the weak scale, in general, we still have
$Q>\xi $ and then the form of the coupling is not modified when one
works with a cut-off scale much below the Planck scale, see also the
discussion after Eq.~(\ref{potqphi2}).

\par

We now discuss the constraint on the parameter characterizing the
inflaton sector, \ie the mass $m$ of the field. In order to simplify
the discussion, we will assume that the initial conditions are such
that quintessence field is always subdominant.  In this situation the
quantum fluctuations of the inflaton field are at the origin of the
CMB anisotropy observed today. As a consequence, and as is well-known,
the COBE and WMAP normalizations fix the coupling constant of the
inflaton potential, namely the mass $m$ in the present context. More
precisely, for small $\ell $, the multipole moments are given by
\begin{equation}
C_{\ell }=\frac{2H^2_{\rm inf}}{25\epsilon \mP^2}\frac{1}{\ell (\ell
+1)}\,
\end{equation}
and what has been actually measured by the COBE and WMAP satellites is
$Q^2_{\rm rms-PS}/T^2=5C_2/(4\pi )\simeq \left(18\times
10^{-6}/2.7\right)^2\simeq 36\times 10^{-12}$. The quantity $H_{\rm
inf}$ is the Hubble parameter during inflation and is related to the
potential by the slow-roll equation $H_{\rm inf}^2\simeq \kappa V_{\rm
inf}/3$ evaluated at Hubble radius crossing. Putting everything
together, we find that the inflaton mass is given by
\begin{equation}
\left(\frac{m}{\mP}\right)^2\simeq 45 \pi \left(N_*+\frac12
\right)^{-2} \frac{Q^2_{\rm rms-PS}}{T^2}\, ,
\end{equation}
that is to say
\begin{equation}
m\simeq 1.3\times 10^{-6}\times \mP\, .
\end{equation}
All the parameters of the potential are now specified.

\par

Let us now discuss in more detail what are the conditions under which
the inflaton field is always dominant. The quintessence energy density
must be smaller than the inflaton energy density. This gives a lower
bound on the possible values of the field $Q$ which can be expressed
as
\begin{equation}
\frac{Q_{\rm low}}{\mP} \simeq \left(\frac{m}{H_0}\frac{\phi
}{\mP}\right)^{-1/p}\simeq 10^{-55/p}\left(\frac{\phi
}{\mP}\right)^{-1/p}\, ,
\end{equation}
where we have used the value of $m\simeq 10^{-6}\times \mP$ obtained
before. The fact that we obtain a lower bound is consistent with the
fact that the potential is an inverse power-law of the quintessence
field: the smaller the field is, the larger the corresponding energy
density is. Secondly, there exists also an upper bound coming from the
fact that the interaction energy density must be smaller than the
inflaton energy density. Concretely, this gives
\begin{eqnarray}
\label{qup}
\frac{Q_{\rm up}}{\mP} &\simeq &\sqrt{\frac{1}{4\pi ^2}\left(\frac{\phi
}{\mP}\right)^{-2}-2\left(\frac{\xi }{\mP}\right)^2}
\nonumber \\
&\simeq & \left(\frac{\phi }{\mP}\right)^{-1}\, ,
\end{eqnarray}
where we have used the fact that the maximal value of the inflaton
field is $\phi \simeq 10^6\times \mP$, see below.

\par

Under the condition that $Q_{\rm low}<Q<Q_{\rm up}$, the behavior of
the background is determined by the energy density of the inflaton and
it is well-known that, in this case, the slow-roll approximation is
valid.  The slow-roll approximation is controlled by two parameters
(in fact, at leading order, there are three relevant slow-roll
parameters but we will not need the third one) defined by~\cite{MS}
\begin{equation}
\epsilon \equiv -\frac{\dot{H}}{H^2}\, ,\quad
\delta =-\frac{\dot{\epsilon}}{2H\epsilon }+\epsilon \, .
\end{equation}
In the present context, where the inflaton potential is proportional
to $m^2\phi ^2$, the slow-roll parameters are given by
\begin{equation}
\epsilon = \frac{1}{2N_*+1}, \quad
\delta = 0\, ,
\end{equation}
where $N_*\simeq 60$ is the number of e-folds between the time at
which scales of astrophysical interest today left the Hubble radius
during inflation and the end of inflation. In the situation where
these parameters are small, namely $\epsilon \ll 1$ and $\delta \ll
1$, the equation of motion of the inflaton field can be easily
integrated. For this purpose, it is convenient to express everything
in terms of the number of e-folds (not to be confused with $N_*$)
defined by
\begin{equation}
N\equiv \ln \left(\frac{a}{a_{\rm ini}}\right)\, ,
\end{equation}
such that, at the beginning of inflation, one has $N=0$. Then, in the
slow-roll approximation, one obtains that the evolution of the field
is given by
\begin{equation}
\frac{\phi }{\mP}=\sqrt{\left(\frac{\phi _{\rm
ini}}{\mP}\right)^2 -\frac{N}{2\pi }}\, ,
\end{equation}
where $\phi _{\rm ini}$ is the initial value of the field. This value
is related to the total number of e-folds given by
\begin{equation}
N_{_{\rm T}}=2\pi \left(\frac{\phi _{\rm ini}}{m_{_{\rm
Pl}}}\right)^2 -\frac{1}{2}\, .
\end{equation}
If $N_{\rm min}$ is the minimum number of e-folds required in order to
solve the problems of the hot big-bang model ($N_{\rm min}\simeq 60$)
then one has
\begin{equation}
\phi _{\rm ini}>m_{_{\rm Pl}}\sqrt{\frac{1}{2\pi }\left(N_{\rm min}
+\frac{1}{2}\right)}\simeq 3.1\times \mP\, .
\end{equation}
There exists also an upper bound for the value of the inflaton field
which corresponds to the situation where the potential energy density
$m^2\phi ^2/2$ is Planckian. Using that $m\simeq 10^{-6}\times \mP$
this immediately gives that $\phi _{\rm max}\simeq 10^{6}\times
\mP$. This is the value of $\phi _{\rm max}$ that we considered
before.

\subsection{Analytical Study of the Klein-Gordon equation}

We now turn to the resolution of the quintessence equation of motion,
\ie the Klein-Gordon equation. It can be written as
\begin{eqnarray}
\ddot{Q}+3H\dot{Q}+\frac{\partial }{\partial Q}
V\left(\phi ,Q\right)=0 \, .
\end{eqnarray}
with $H\equiv \dot{a}/a$ is the Hubble parameter which only depends on
the inflaton energy density. A dot denotes a derivative with respect
to cosmic time. We now work with the new time variable introduced
before, namely the number of e-folds. One gets
\begin{eqnarray}
\label{kgefold}
& &\frac{{\rm d}^2}{{\rm d}N^2}\left(\frac{Q}{\mP}\right)
+\left(3+\frac{1}{H}\frac{{\rm d}H}{{\rm d}N}\right)
\frac{{\rm d}}{{\rm d}N}\left(\frac{Q}{\mP}\right)
\nonumber \\
& & +\left(\frac{\mP }{H}\right)^2
\frac{\partial f}{\partial (Q/\mP)}=0\, .
\end{eqnarray}
The main feature of the above equation is that the potential is now
explicitly time-dependent because of the interaction of $Q$ with the
inflaton, \ie we have $f\left(\phi ,Q\right)=f\left(N,Q\right)$. This
renders this equation very difficult to solve exactly.  Therefore, in
order to get some analytical approximate solution, it is necessary to
make some assumptions. With a very good accuracy, the fact that
$Q_{\rm low}<Q<Q_{\rm up}$ implies that the potential contains two
dominant terms and can be approximated as $\sim Q^{-2p}+Q^2$. As a
consequence, its derivative can expressed as
\begin{eqnarray}
\label{derpot}
\frac{\partial f}{\partial (Q/\mP)} &\simeq &
-\frac{p}{4}\left(\frac{M}{\mP}\right)^{4+2p}
\left(\frac{Q}{\mP}\right)^{-2p-1}
\nonumber \\
& & +4 \pi
^2\left(\frac{m}{\mP}\right)^2 \frac{Q}{\mP} \left(\frac{\phi
}{\mP}\right)^4 \, .
\end{eqnarray}
This means that the quintessence field evolves in a time-dependent
potential which possesses a minimum. Of course, this minimum is
explicitly time-dependent and can be expressed as
\begin{widetext}
\begin{equation}
\label{Qmin}
Q_{\rm min}(N)=\mP \times \left\{\frac{p}{16 \pi
^2}\left(\frac{H_0}{\mP}\right)^2 \left(\frac{m}{\mP}\right)^{-2}
\left[\frac{\phi (N)}{\mP}\right]^{-4} \right\}^{1/[2(p+1)]}\, .
\end{equation}
\end{widetext}
As shown below, this equation turns out to be one of the main result
of the present article. Indeed, we will demonstrate that, after a
period of rapid oscillations, the quintessence field always tends
toward the above solution. Therefore, in the case where the
interaction between the inflaton and the quintessence field is
important, $Q_{\rm min}$ can be viewed as a kind of attractor solution
since, regardless of the initial conditions, the final value of the
field is always given by $Q_{_{\rm min}}$.

\par

Several remarks are in order at this stage. Firstly, let us evaluate
the typical time of evolution of the minimum. It is given by $\Delta
N\simeq Q_{\rm min}/({\rm d}Q_{\rm min}/{\rm d}N)=[{\rm d}\ln Q_{\rm
min}/{\rm d}N]^{-1}$. From Eq.~(\ref{Qmin}), one has $Q_{\rm
min}\propto H^{-2/(p+1)}$. Therefore, this implies that
\begin{equation}
\Delta N_{\rm min} \simeq \left \vert \frac{p+1}{2\epsilon
}\right\vert\gg 1\, ,
\end{equation}
where $\epsilon =-({\rm d}H/{\rm d}N)/H$ is the first slow-roll
parameter.

\par

Secondly, it is interesting to calculate the effective mass of the
quintessence field at the minimum of its time-dependent
potential. Using Eq.~(\ref{Qmin}), one obtains
\begin{equation}
\frac{m_{\rm eff}^2}{\mP^2}=\frac{\partial f^2}{\partial
(Q/\mP)^2}\biggl\vert _{\rm min} \simeq 8\pi
^2(p+1)\left(\frac{m}{\mP}\right)^{2} \left[\frac{\phi
(N)}{\mP}\right]^{4}\, .
\end{equation}
Therefore, one has
\begin{equation}
\frac{m_{\rm eff}^2}{H^2}=6\pi (p+1)\left[\frac{\phi
(N)}{\mP}\right]^{2}>1 \, ,
\end{equation}
and we conclude that, at its minimum, the quintessence field is not a
light field.

\par

Thirdly, we are now in a position where one can study how small
fluctuations behave around the time-dependent minimum. The
fluctuations are given by $\delta Q\equiv Q-Q_{\rm min}$ and their
evolution is governed by the equation
\begin{eqnarray}
\frac{{\rm d}^2}{{\rm d}N^2}\left(\frac{\delta Q}{\mP}\right)
+3\frac{{\rm d}}{{\rm d}N}\left(\frac{\delta Q}{\mP}\right)
+\left(\frac{m_{\rm eff}}{H}\right)^2
\frac{\delta Q}{\mP}=0\, ,
\end{eqnarray}
where, in the damping term, we have neglected the derivative of the
Hubble parameter which is nothing but the slow-roll parameter
$\epsilon $. Using the expression of the effective mass established
before and the expression of the inflaton in the slow-roll
approximation, one finds that the solution can be expressed as
\begin{equation}
\label{soldeltaQ}
\frac{\delta Q}{\mP}={\rm e}^{-3N/2}
\left[A_1{\rm Ai}\left(-x\right)+A_2{\rm Bi}\left(-x\right)\right],
\end{equation}
where $A_1$ and $A_2$ are two constants determined by the initial
conditions. The functions Ai and Bi are the Airy functions~\cite{Grad}
\footnote{Another method to solve the Klein-Gordon equation, under the
assumption that the slow-roll hypothesis is valid for the inflaton, is
the following. Instead of directly neglecting $\epsilon =-({\rm
d}H/{\rm d}N)/H$ in the damping term of Eq.~(\ref{kgefold}) as we did
before, one works with the dimensionless field $q(N)$ defined by
\begin{equation}
\frac{Q}{\mP} \equiv g(N)q(N)\equiv
\left(\frac{H}{\mP}\right)^{-1/2}{\rm e}^{-3N/2}q(N)\, .
\end{equation}
Then, from Eq.~(\ref{kgefold}), it is easy to show that the field
$q(N)$ obeys
\begin{eqnarray}
& & \frac{{\rm d}^2q}{{\rm d}N^2}+\left[-\frac{1}{2H}\frac{{\rm
d}^2H}{{\rm d}N^2}+\frac{1}{4H^2}\left(\frac{{\rm d}H}{{\rm
d}N}\right)^2-\frac{15 }{4H}\frac{{\rm d}H}{{\rm d}N} -\frac94\right]q
\nonumber \\ & & +\frac{1}{g(N)}\left(\frac{H}{\mP}\right)^{-2}
\frac{\partial f}{\partial (Q/\mP)}\biggl\vert _{Q/\mP=g(N)q(N)}=0\, .
\end{eqnarray}
In the second term between squared brackets, the various derivatives
of the Hubble parameter can be expressed in terms of the slow-roll
parameters. Explicitly, this term reads $\epsilon (3\epsilon +2\delta
)/2+\epsilon ^2/4 +15\epsilon /4 -9/4\simeq -9/4$. Therefore, the
Klein-Gordon equation can be simplified further and we obtain
\begin{eqnarray}
\frac{{\rm d}^2q}{{\rm d}N^2}-\frac94q +
\frac{1}{g(N)}\left(\frac{H}{\mP}\right)^{-2} \frac{\partial
f}{\partial (Q/\mP)}\biggl\vert _{g(N)q(N)}\simeq 0\, .
\end{eqnarray}
Using the fact that the potential is given by $m_{\rm eff}^2\phi
^2/2$, hence we are now studying $\delta Q\equiv g(N)q(N)$, the above
equation takes the form
\begin{eqnarray}
\frac{{\rm d}^2q}{{\rm d}x^2}+xq=0 \, ,
\end{eqnarray}
where $x$ is defined in Eq.~(\ref{defx}). As before, this equation can
be solved in terms of Airy functions and this gives
\begin{equation}
\frac{\delta Q}{\mP}=\left(\frac{H}{\mP}\right)^{-1/2}{\rm e}^{-3N/2}
\left[B_1{\rm Ai}\left(-x\right)+B_2{\rm Bi}\left(-x\right)\right]\, .
\end{equation}
This solution should be compared with Eq.~(\ref{soldeltaQ}). We see
that the equation are similar up to the factor $(H/\mP)^{-1/2}$. As
our approximation is valid during a few e-folds only, the Hubble
parameter can be considered as a constant and then the two solutions
are identical.} and the quantity $x$ is defined by
\begin{equation}
\label{defx}
x\equiv 3^{-2/3}(p+1)^{-2/3}\left\{6\pi (p+1)\left[\frac{\phi
(N)}{\mP}\right]^{2}-\frac94\right\}\, .
\end{equation}
Initially, and during a few e-foldings, one has $x\gg 1$. In this
case, one can use the asymptotic expression of the Airy
function~\cite{Grad} and one obtains
\begin{eqnarray}
\frac{\delta Q}{\mP} &\simeq & {\rm e}^{-3N/2}\pi ^{-1/2}x^{-1/4}
\biggl[A_1\sin \left(\frac23 x^{3/2}+\frac{\pi }{4}\right)
\nonumber \\
& & +A_2\cos \left(\frac23 x^{3/2}+\frac{\pi }{4}\right)\biggr]\, .
\end{eqnarray}
{}From this expression, one sees that one has damped oscillations. The
period of the oscillations can be very easily estimated. One has
\begin{equation}
\label{DeltaNosci}
\Delta N_{\rm osci} \simeq \frac{2\pi }{\sqrt{3(p+1)}}N_{_{\rm
T}}^{-1/2}\, ,
\end{equation}
where we recall that $N_{_{\rm T}}$ is the total number of e-folds
during inflation. For a typical model with $p=3$ and $N_{_{\rm
T}}\simeq 60$, one gets $\Delta N_{\rm osci}\simeq 0.24$. The previous
equation also means that if the inflaton field starts at large values,
then the period of the oscillations will be extremely rapid. For
instance, if inflation starts at Planckian density, then $\phi _{\rm
ini}\simeq 10^6\mP$ which implies that $N_{_{\rm T}}\simeq
10^{12}$. As a consequence, one can get values as small as $\Delta
N_{\rm osci}\simeq 10^{-6}$.

\par

Therefore, from the above considerations, one reaches the conclusion
that
\begin{equation}
\frac{\Delta N_{\rm min}}{\Delta N_{\rm osci}}={\cal
O}(1)\frac{\sqrt{N_{_{\rm T}}}}{\epsilon}\gg 1 \, .
\end{equation}
This means that the oscillatory phase is very quick in comparison with
the typical time scale of evolution of the minimum. To put it
differently, the minimum can be considered as motionless or as
``adiabatic'' as the field rapidly oscillates and quickly joins its
minimum. Therefore, $Q_{\rm min}(N)$ can be viewed as an attractor
since it does not depend on the initial conditions for the
quintessence field.

\par

We have established the above result under the assumption that the
initial deviation from the attractor $Q_{\rm min}$ is not too large
(or, in other words, that $\delta Q$ is not too large). What happens
if this is not the case, \ie if $\vert Q_{\rm ini}-Q_{\rm min}\vert
\gg 1$ or $\delta Q_{\rm ini}\gg 1$? Is the attractor still joined
rapidly enough (\ie before the end of inflation)? {\it A priori}, to
answer this question requires a full integration of the equation of
motion (or a numerical integration, see the next subsection) which is
not possible. However, we can gain some partial insights using the
following considerations. If one has $Q_{\rm ini}\ll Q_{\rm min}$ then
the term proportional to $Q^{-2p}$ dominates in the potential and the
Klein-Gordon equation remains non linear hence difficult to
integrate. But if we now assume that we start from a situation where
$Q_{\rm ini}\gg Q_{\rm min}$, then the term proportional to $Q^2$
dominates in the potential. As a consequence, the derivative of the
potential can be written as, see Eq.~(\ref{derpot})
\begin{eqnarray}
\frac{\partial f}{\partial (Q/\mP)} &\simeq & 4 \pi
 ^2\left(\frac{m}{\mP}\right)^2 \left(\frac{\phi
 }{\mP}\right)^4 \frac{Q}{\mP}\, .
\end{eqnarray}
Therefore, the Klein-Gordon equation is now linear and can be
integrated. In fact, one obtains the same potential as before for
$\delta Q$, up to an unimportant factor $2(p+1)$, except that now
$Q-Q_{\rm min}>0$ needs not to be small. As a consequence the solution
will read the same, that is to say, roughly speaking, $Q\simeq Q_{\rm
ini}\exp(-3N/2)$, this solution being valid provided $Q\gg Q_{\rm
min}$.

\par

Equipped with this solution, one can now estimate how many e-folds are
necessary for the field to roll down the potential from a given
initial condition and to reach the region of the minimum. When the
field enters this region, the previous solution is no longer valid
because the term $Q^{-2p}$ starts playing a role but, on the other
hand, since the field is now close the $Q_{\rm min}$ the calculation
of $\delta Q$ applies. In order to get an upper bound on the number
e-folds, let us assume that $Q$ is initially as far as possible from
the minimum, \ie $Q_{\rm ini}=Q_{\rm up}$, see Eq.~(\ref{qup}). Then
the number of e-folds $N$ is solution of the algebraic equation
\begin{equation}
\frac{Q_{\rm min}(N)}{\mP}=\left(\frac{\phi _{\rm
ini}}{\mP}\right)^{-1} {\rm e}^{-3\Delta N/2}\, .
\end{equation}
The solution can be expressed in terms of the Lambert function
$W_0$~\cite{lambert} defined by the relation
$W(z)\exp[W(z)]=z$. Explicitly, one obtains
\begin{widetext}
\begin{equation}
\Delta N=-(p+1)N_{_{\rm T}}+\frac23W_0\left\{\frac32(p+1)N_{_{\rm T}}
\left(\frac{\phi _{\rm ini}}{\mP}\right)^{(3+p)/(1+p)}
\left(\frac{p}{16\pi ^2}
\frac{H_0^2}{\mP^2}\frac{\mP^2}{m^2}\right)^{-1/[2(p+1)]}
{\rm e}^{3(p+1)N_{_{\rm T}}/2}\right\}\, .
\end{equation}
\end{widetext}
Since the argument of the Lambert function $W_0$ is very large, one
can use the approximation $W_0(z)\simeq \ln (z)$. In this case, one
gets
\begin{eqnarray}
& & \Delta N \sim \frac23\ln \left[\frac32(p+1)N_{_{\rm T}}\right]
+\frac{2(3+p)}{3(1+p)}\ln \left(\frac{\phi _{\rm ini}}{\mP}\right)
\nonumber \\
&- &\frac{2}{3(p+1)}\ln \left(\frac{H_0}{\mP}\right)
+\frac{2}{3(p+1)}\ln \left(\frac{m}{\mP}\right)\, .
\end{eqnarray}
For the fiducial model with $p=3$, $\phi _{\rm ini}=3.1\mP$ and
$N_{_{\rm T}}=60$ one obtains $\Delta N\simeq 26<60$. Therefore, even
in the extreme case where the quintessence field starts at $Q_{\rm
up}$, inflation lasts enough e-folds so that $Q$ has time to reach the
attractor.

\par

In conclusion, in this subsection, we have shown that, during
inflation, the evolution of the quintessence field is characterized by
two very different time scales. One scale describes the evolution of
the adiabatic time-dependent minimum while the second one represents
the period of the rapid oscillations around this minimum. We have
demonstrated that $Q_{\min}$ is in fact an attractor and that,
regardless of the initial conditions, the quintessence field always
has enough e-folds during inflation to join this attractor. Although
the above conclusion has been established in the quadratic part of the
potential, it is in fact true even in the regime where the potential
is proportional to $Q^{-2p}$ as confirmed by a numerical study of the
Klein-Gordon equation.

\subsection{Numerical Study of the Klein-Gordon Equation}

We have just seen that the equation of motion cannot be analytically
integrated with the complete potential. As a consequence, the part
where the potential is proportional to $Q^{-2p}$ has not been explored
for values of the initial conditions far from the minimum. In this
subsection, we perform the integration numerically. The difficulty is
that we have to deal with very small quantities. It is therefore
necessary to absorb these small quantities into a redefinition of the
quintessence field which greatly facilitates the numerical
integration. For this purpose, we write
\begin{equation}
\frac{Q}{\mP}=\lambda {\cal Q}\, ,
\end{equation}
where $\lambda $ is a constant. It is easy to show that, if $\lambda $
is chosen to be
\begin{equation}
\lambda
=\left(\frac{m}{\mP}\right)^{-1/(p+1)}\left(\frac{H_0}{\mP}\right)^{1/(p+1)}\,
,
\end{equation}
then we can remove the dangerous coefficients from the equation of motion
which now reads
\begin{eqnarray}
& &\frac{{\rm d}^2{\cal Q}}{{\rm d}N^2} +\left(3+\frac{1}{H}\frac{{\rm
d}H}{{\rm d}N}\right) \frac{{\rm d}{\cal Q}}{{\rm d}N} +\frac{3}{4\pi}
\left(\frac{\phi }{\mP}\right)^{-2}
\nonumber \\
& &\times
\left[-\frac{p}{4} {\cal Q}^{-2p-1}
+4\pi^2\left(\frac{\phi }{\mP}\right)^4{\cal Q}\right]=0 \, .
\end{eqnarray}
In particular, it is interesting to evaluate how the time-dependent
minimum looks like after the rescaling. From Eq.~(\ref{Qmin}), one
gets
\begin{equation}
{\cal Q}_{\rm min}(N)=\left(\frac{p}{16\pi ^2}\right)^{1/(2p+2)}\times
\left(\frac{\phi }{\mP}\right)^{-2/(p+1)}\, .
\end{equation}
\begin{figure*}
\includegraphics[width=.85\textwidth,height=.65\textwidth]{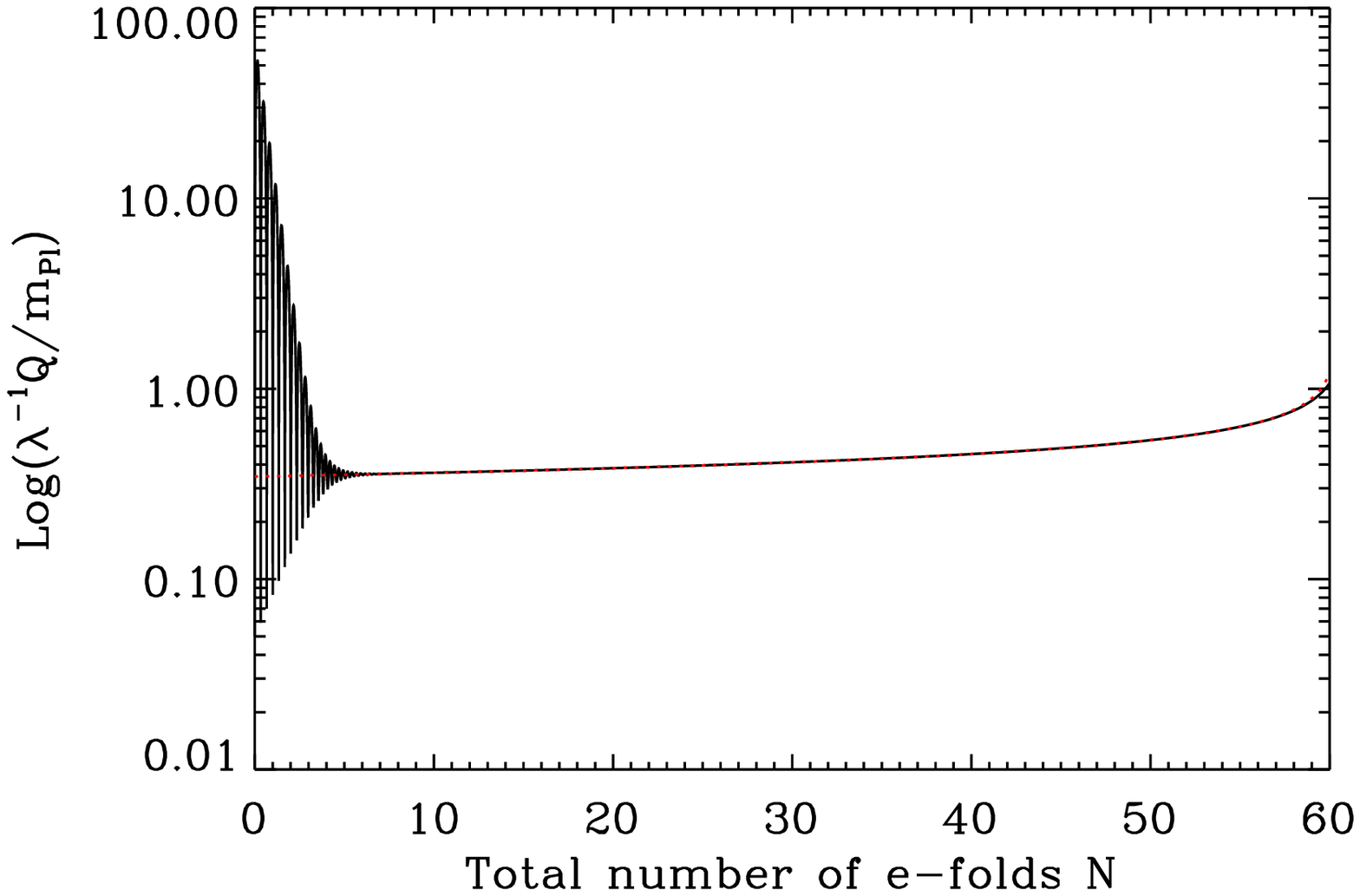}
\includegraphics[width=.85\textwidth,height=.65\textwidth]{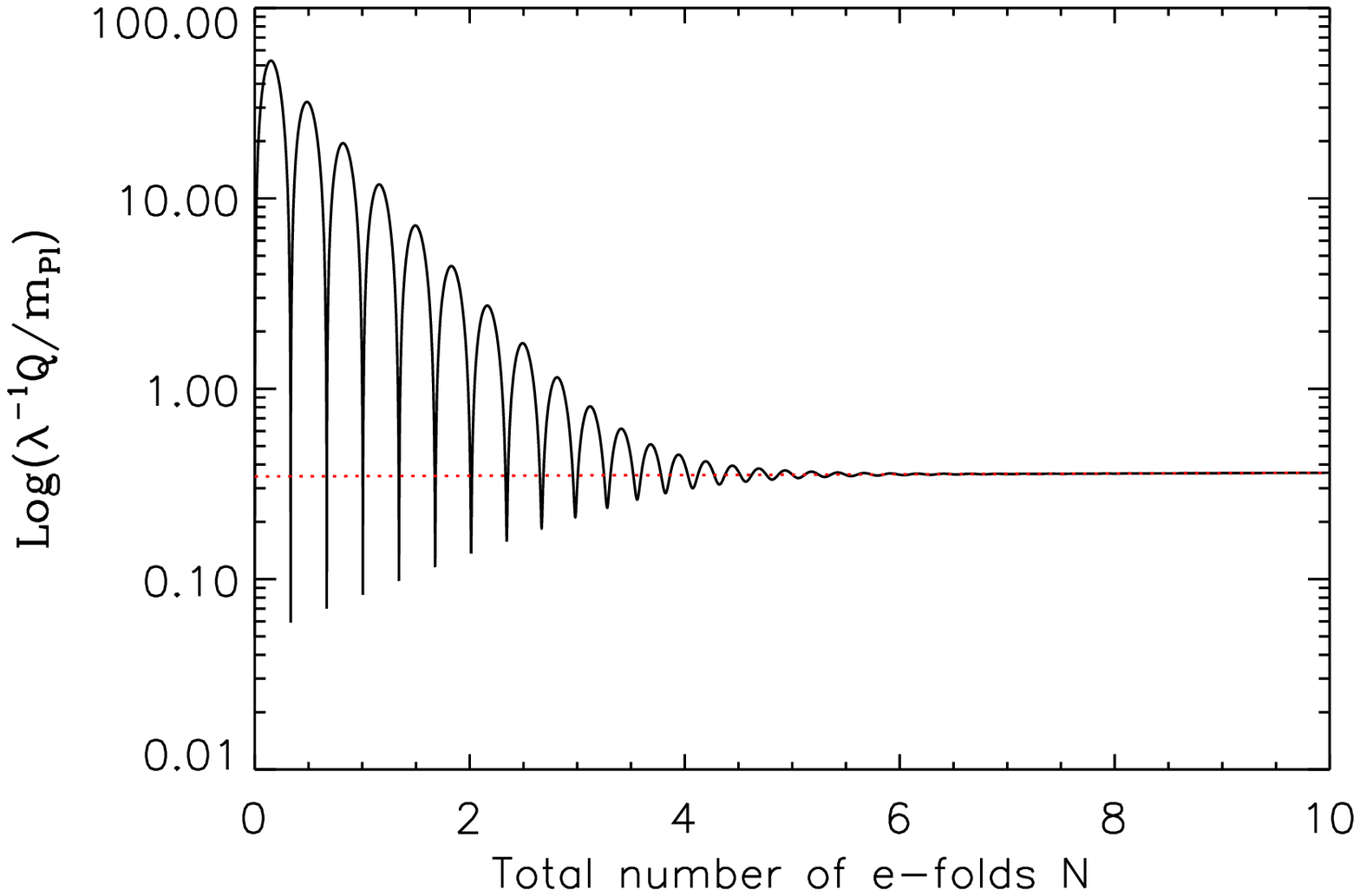}
\caption{Upper panel: Evolution of the quintessence field during
inflation (solid black line). The model of inflation is chaotic
inflation with a massive potential and the initial value of the
inflaton is chosen to be $\phi_{\rm ini}=3.1\times \mP$ corresponding
to a total number of e-folds $N_{_{\rm T}}=60$. The potential of the
quintessence field is of the Ratra-Peebles type with $p=3$. The
initial value of the quintessence field is taken to be ${\cal Q}_{\rm
ini}=0.05$ or $Q_{\rm ini}\simeq 8.3\times 10^{-16}\mP$. This initial
value is such that $Q_{\rm ini}<Q_{\rm min}$ where $Q_{\rm min}$ is
the time-dependent minimum of the effective potential. The evolution
of $Q_{\rm min}(N)$ is given by the dotted red curve. Bottom panel: a
zoom of the upper figure at the beginning of inflation.}
\label{QN_0.05}
\end{figure*}
\begin{figure*}
\includegraphics[width=.85\textwidth,height=.65\textwidth]{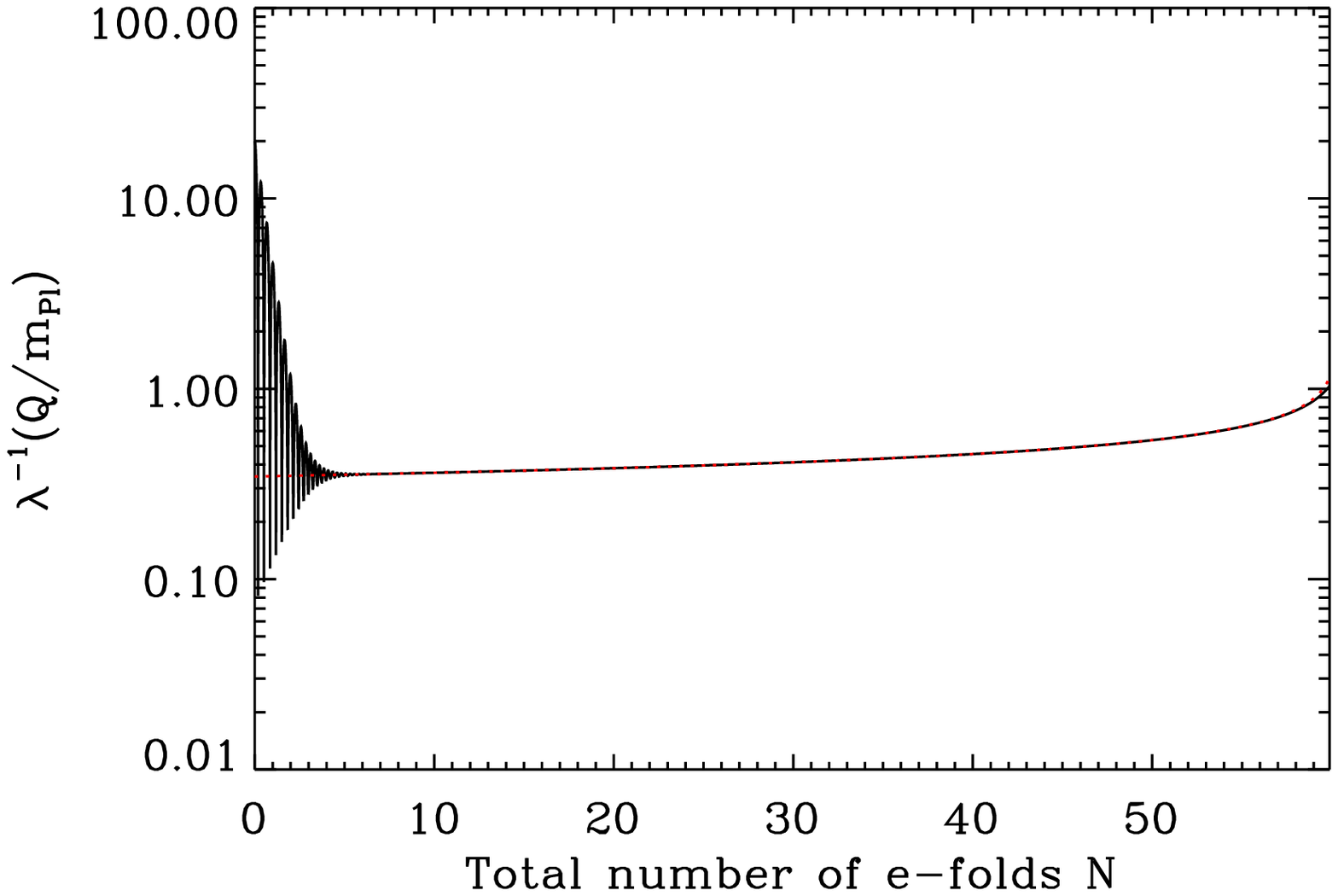}
\includegraphics[width=.85\textwidth,height=.65\textwidth]{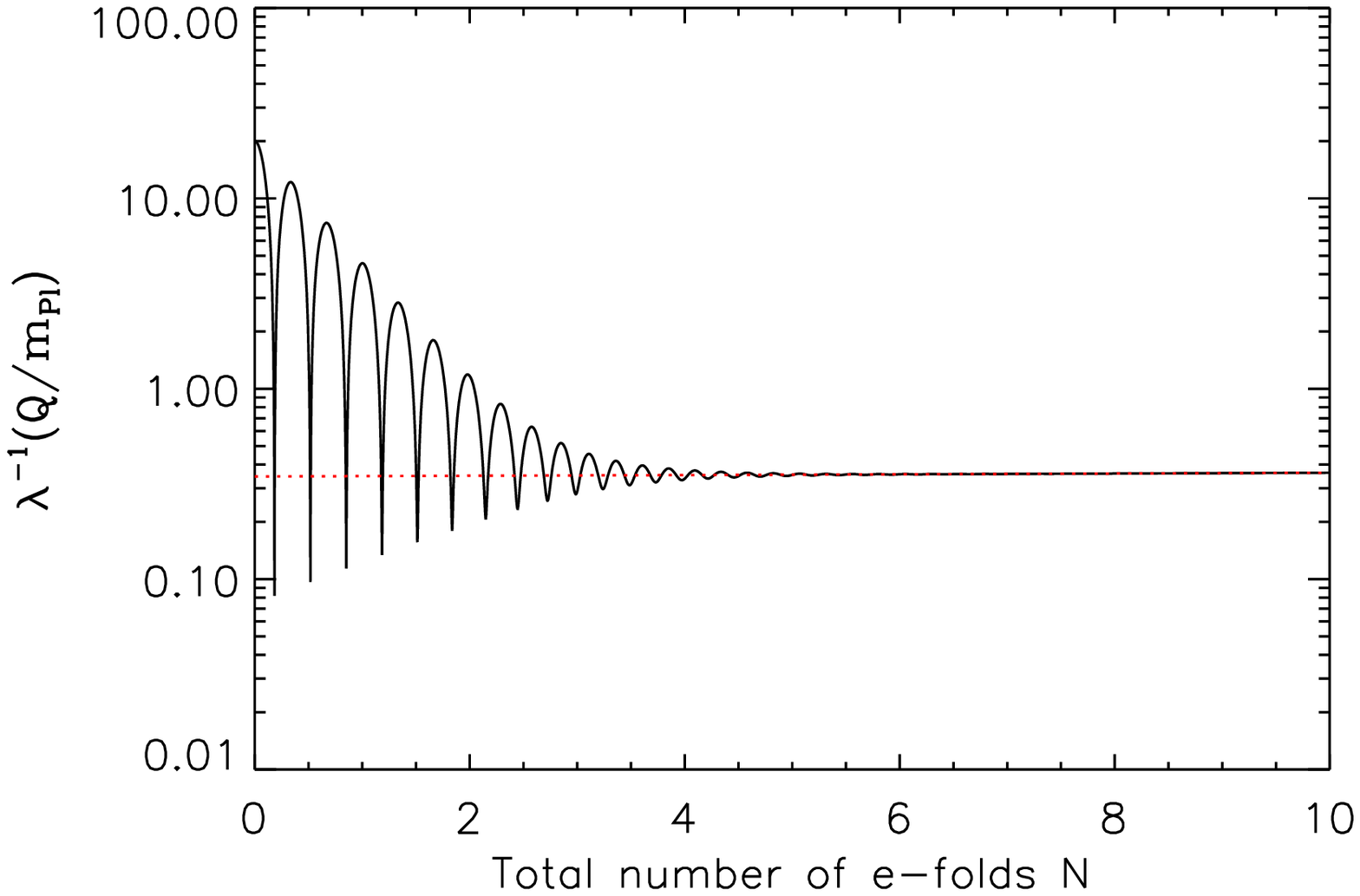}
\caption{Upper panel: Evolution of the quintessence field during
inflation (solid black line) with the initial condition ${\cal Q}_{\rm
ini}=20$ or $Q_{\rm ini}\simeq 3.3\times 10^{-13}\mP$. This initial
condition corresponds to a situation where $Q_{\rm ini}>Q_{\rm
min}$. The parameters characterizing the model are identical to those
used in Fig.~\ref{QN_0.05}. The dotted red curve represents the
time-dependent minimum $Q_{\rm min}(N)$. Bottom panel: a zoom of the
upper figure at the beginning of inflation.}
\label{QN_20}
\end{figure*}

The results of the numerical integration are presented in
Figs.~\ref{QN_0.05} and~\ref{QN_20} for two different initial
conditions. We always assume that the inflaton field starts at $\phi
_{\rm ini}=3.1\mP$ which, as already mentioned, means that the total
number of e-folds during inflation is $N_{_{\rm T}}=60$. The
quintessence potential has been chosen such that $p=3$. This also
completely specifies ${\cal Q}_{\rm min}$ and in particular we have
${\cal Q}_{\rm min}(N=0)\simeq 0.34$ as can be checked directly on the
figures. The evolution of ${\cal Q}_{\rm min}(N)$ is represented by
the red dotted curve in Figs.~\ref{QN_0.05} and~\ref{QN_20}. Moreover,
with the values of $H_0$ and $m$ discussed before, the rescaling
constant $\lambda $ is equal to $\lambda \simeq 1.6\times 10^{-14}$
and, therefore, the initial value of the attractor is in fact $Q_{\rm
min}(N=0)\simeq 5.6\times 10^{-15}\mP$.

\par

In Fig.~\ref{QN_0.05}, one has ${\cal Q}_{\rm ini}=0.05$ or $Q_{\rm
ini}\simeq 8.3\times 10^{-16}\mP$. Initially, the field is therefore
in the region where the potential is proportional to $Q^{-2p}$, \ie
the region which was explored analytically before.  We see that the
evolution is very similar to what was discussed before. We have a
period of rapid oscillations and then, after a few e-folds, the
attractor is joined. We notice that the period of these oscillations
is in full agreement with the estimate of Eq.~(\ref{DeltaNosci}). One
can check that the amplitude of the oscillations decreases as
$\exp(-3N/2)$ as demonstrated in the previous subsection. We conclude
that all the properties established before are confirmed by the
numerical study, even in the part of the potential where it is
proportional to $Q^{-2p}$.  However, one should also notice that, if
the field is initially very displaced from its minimum such that
$Q_{\rm ini}\ll Q_{\rm min}$, then the simple Fortran code used to
integrate the equation of motion can quickly run into numerical
problems. This is probably due to the fact that $Q^{-2p}$ is a very
steep potential.  Despite this remark, one sees no reason why, in this
regime, the evolution of $Q$ should be different from what has been
described before.

\par

In Fig.~\ref{QN_20}, one has ${\cal Q}_{\rm ini}=20$ or $Q_{\rm
ini}\simeq 3.3\times 10^{-13}\mP$. This time, one starts from the
other part of the potential, where $Q_{\rm in}>Q_{\rm min}$ and
$V(Q)\propto Q^2$. The remarks made before also apply to this case
which appears to be in full agreement with the analytical estimates of
the previous subsection.

\section{Discussion and Conclusions}

In order to study the influence of the interaction term and to compare
its effect with the standard case, it is interesting to give the
evolution of the quintessence field when this one does not interact
with the inflaton (and when $Q$ remains a test field). In particular,
we are interested in calculating, for a given initial condition at the
beginning of inflation, the value of $Q$ at the end of inflation (or
at the beginning of the radiation dominated era) in both cases (\ie
with and without interaction). The case without interaction can be
easily treated because the Klein-Gordon equation can be integrated in
the slow-roll approximation, see Ref.~\cite{JMarcello}. In fact, this
equation can be re-written as
\begin{equation}
\frac{\ddot Q}{H(\phi)\dot Q}= -\frac{V_{_{\rm
RP}}''(Q)}{3H^2(\phi)}+\epsilon \, ,
\end{equation}
where a dot denotes a derivative with respect to cosmic time. In the
above equation $V_{_{\rm RP}}(Q)$ now means the Ratra-Peebles
potential, namely $V_{_{\rm RP}}(Q)\propto Q^{-2p}$ since this is the
potential for the quintessence field in absence of any interaction
with the inflaton field. Due to the smallness of the parameter
$\epsilon $, the slow roll approximation can be applied to the
equations describing the motion of the quintessence field without
interaction if the following condition is satisfied
\begin{equation}
\frac{V_{_{\rm RP}}''(Q)}{3H^2(\phi)} \ll 1 \, .
\end{equation}
If one applies this condition to the Ratra-Peebles potential, one
gets
\begin{eqnarray}
\label{slowQ}
\left(\frac{Q}{\mP}\right)^{2(p+1)} &\gg &\frac{p(2p+1)}{2\pi }
\left(\frac{M}{\mP}\right)^{4+2p}
\nonumber \\
& & \times \left(\frac{m}{\mP}\right)^{-2}
\left(\frac{\phi}{\mP}\right)^{-2}\, .
\end{eqnarray}
This formula is similar to Eq.~(48) of Ref.~\cite{JMarcello}. It can
also be re-written as
\begin{equation}
\label{srcond}
\lambda ^{-1}\frac{Q}{\mP}> {\cal
F}(p)\left(\frac{\phi}{\mP}\right)^{-1/(p+1)}\, ,
\end{equation}
where ${\cal F}(p)=[p(2p+1)/(2\pi )]^{1/[2(p+1)]}\simeq {\cal O}(1)$.
Then, integrating the Klein-Gordon equation leads to the following
expression
\begin{widetext}
\begin{eqnarray}
Q_{\rm no \, inter}(N) &=& Q_{\rm
ini}\left[1-4p(p+1)\left(\frac{m}{\mP}\right)^{-2}
\left(\frac{M}{\mP}\right)^{4+2p} \left(\frac{Q_{\rm
ini}}{\mP}\right)^{-2(p+1)} \ln\frac{\phi (N)}{\phi _{\rm
ini}}\right]^{1/2(p+1)}\, ,
\\
&\simeq & Q_{\rm ini}\left[1+2p(p+1)\left(\frac{m}{\mP}\right)^{-2}
\left(\frac{M}{\mP}\right)^{4+2p} \left(\frac{Q_{\rm
ini}}{\mP}\right)^{-2(p+1)} \frac{N}{N_{_{\rm T}}}\right]^{1/2(p+1)}\,
,
\end{eqnarray}
\end{widetext}
where in the last equality we have used the fact that the evolution of
the inflaton field can be approximated by $\phi \simeq \phi _{\rm
ini}[1-N/(2N_{_{\rm T}})]$. The subscript ``no inter'' just reminds
that the above equation gives $Q$ in the case where there is no
interaction between $Q$ and $\phi $. From this expression, we deduce
that the quintessence field is frozen if
\begin{equation}
\label{frozen}
\lambda ^{-1}\frac{Q_{\rm ini}}{\mP}\gsim 1\, .
\end{equation}
If this condition is satisfied, then obviously the
condition~(\ref{srcond}) is also satisfied. The contrary is not
necessarily true but, as show for instance in Fig.~3 of
Ref.~\cite{JMarcello}, this only concerns a small range of initial
conditions. Therefore, we can consider that $Q_{\rm no \, inter}\simeq
Q_{\rm ini}$.

\par

We are now a position where the values of the quintessence field with
and without interaction can be compared at the end of
inflation. Inflation stops when the slow-roll parameter $\epsilon $ is
equal to unity corresponding to $\phi _{\rm end}/\mP=1/(2\sqrt{\pi
})$. With the interaction term taken into account, the field will be
on the attractor $Q_{\rm min}$ and, therefore, its value at the end of
inflation is just the value of $Q_{\rm min}$ at the end of inflation,
namely
\begin{eqnarray}
\frac{Q_{\rm min}}{\mP}\biggl \vert _{N=N_{_{\rm T}}}&=&\lambda
\left(\frac{p}{16\pi ^2}\right)^{1/(2p+2)}\times
\left(\frac{1}{2\sqrt{\pi}} \right)^{-2/(p+1)}
\\
&=& {\cal H}(p)\lambda \, .
\end{eqnarray}
As already mentioned, the striking feature of this expression is that
it does not depend on $Q_{\rm ini}$. For orders of magnitude estimate,
one can consider that ${\cal H}(p)={\cal O}(1)$. For $p=3$, which is
our fiducial model, one has $Q_{\rm min}(N=N_{_{\rm T}}) \simeq
1.9\times 10^{-14}\mP$. Therefore the ratio of $Q$ at the end of
inflation without the interaction term taken into account to $Q$ at
the end of inflation with the interaction term into account is given
by
\begin{eqnarray}
\frac{Q_{\rm no \, inter}}{Q_{\rm inter}}\biggl \vert _{\rm end}
&\simeq & \frac{Q_{\rm ini}}{\mP}\left(\frac{m}{\mP}\right)^{1/(p+1)}
\left(\frac{H_0}{\mP}\right)^{-1/(p+1)}
\\
&\sim & 10^{55/(p+1)}\times \frac{Q_{\rm ini}}{\mP}\, .
\end{eqnarray}
This ratio is necessarily greater than one, see Eq.~(\ref{frozen}),
and for ``large'' initial conditions can be much bigger than one. This
means that, generically, $Q_{\rm no \, inter}\gg Q_{\rm inter}$, \ie
the effect of the interaction is to force the quintessence field to
remain small during inflation.

\par

Two loopholes could modify the above conclusion. Firstly, we have seen
that the numerical integration, on which our study is based, is valid
only if the initial value of $Q$ is not too far from the initial value
of the minimum. We have used the results obtained under this condition
and have extrapolated them for any initial conditions. If the initial
value of the quintessence field is far from the time-dependent
minimum, we still expect a phase of oscillations. However, since we
have noticed that the amplitude of the oscillations tend to be quite
big even if the initial displacement from the minimum remains
reasonable (this is not surprising for a potential like $Q^{-2p}$
which is very ``abrupt''), it could actually happen that the amplitude
of the oscillations, in the case where the initial displacement is
large, are so big that the assumption that the quintessence field is a
test field becomes violated. This is a regime that has not been
studied in the present article. Some new interesting effects could
occur in this case. However, we suspect that the numerical study of
this situation could be quite tricky.

\par

Secondly, it has been shown in Refs.~\cite{JMarcello} that the quantum
effects could strongly modify the evolution of $Q$ during
inflation. These quantum effects have been calculated in
Refs.~\cite{JMarcello} by means of the stochastic inflation formalism
for a free field. Therefore, what should now be done is to compute
these effects in the case where the interaction term is present. This
is clearly a difficult task which is beyond the scope of the present
paper. In Refs.~\cite{JMarcello}, it has been shown that the quantum
effects can push the quintessence field to quite large values at the
end of inflation. As a consequence, the attractor solution could be
joined only at late time and even after the present time. In this case
the quintessential scenario would lose an attractive feature, namely
its insensitivity to the initial conditions. In this respect, the
results reached in the present article are good news since the effect
of the interaction term seems to retain the field to quite small
values. One could speculate that this could maybe compensate the
influence of the quantum effects.

\par

In conclusion we have shown that the non-renormalizable interactions
between the inflaton and the quintessence field have drastic
consequences during inflation. The quintessence field follows an
attractor and remains small compared to the Planck scale at the end of
inflation. This sets the initial conditions for the quintessence
field.  As is well known, the quintessence field cannot couple
strongly to matter field.  On the contrary non-renormalizable
couplings between the quintessence field and matter, in particular
Cold Dark Matter, may have a crucial impact on the late time physics
of the quintessence field, i.e. on the coincidence problem.  This is
left for future work.

\vspace{0.5cm}
\centerline{\bf Acknowledgments}
\vspace{0.2cm}

We wish to thank C.~Ringeval for many enlightening comments and
discussions.

\end{document}